\documentclass[12pt]{article}
\def\gappeq{\mathrel{\rlap {\raise.5ex\hbox{$>$}}
{\lower.5ex\hbox{$\sim$}}}}

\def\lappeq{\mathrel{\rlap{\raise.5ex\hbox{$<$}}
{\lower.5ex\hbox{$\sim$}}}}

\def\Toprel#1\over#2{\mathrel{\mathop{#2}\limits^{#1}}}

\newcommand{\al}{\alpha}
\newcommand{\ice}[1]{\relax}
\newcommand{\be}{\begin{equation}}
\newcommand{\ee}{\end{equation}}

\newcommand{\g}{\gamma}

\newcommand{\bea}{\begin{eqnarray}}
\newcommand{\eea}{\end{eqnarray}}
\begin{document}
\pagestyle{empty}
\begin{flushright}
\centerline{\normalsize\hfill hep-ph/0103230}
\end{flushright}
\vspace*{5mm}
\begin{center}

{\Large  \bf Calculation of the ${\rm K}^0$-$\bar {\rm K}^0$ mixing
parameter\\
via the QCD sum rules at finite energies}\footnote{\normalsize
Copy of the
paper  published in {\it Phys. Lett.} {\bf B174} (1986) 104.
Received 8 July 1985; revised manuscript received  11 April 1986.
A couple of minor misprints is corrected. Comments, related to 
a (relatively small)  change of the  result  due to changes in the values 
of incorporated phenomenological parameters, are added.}
\\
\vspace*{1cm}
{\bf K.G. Chetyrkin, A.L. Kataev, A.B. Krasulin and A.A. Pivovarov}\\
\vspace{0.3cm}
Institute for Nuclear Research of the Academy of Sciences of Russian 
Federation,\\
60th October Anniversary prospect, 7a, 117312 Moscow, Russian Federation
\end{center}
\vspace*{2cm}
\begin{center}
{\bf ABSTRACT}
\end{center}
\vspace*{5mm}
\noindent
The QCD finite energy sum rules method is used to show that the
parameter of the $K^0$-$\bar K^0$ mixing $\hat{B}$ is mainly
determined by the value of
$gm_s\langle\overline{d}G_{\mu\nu}\sigma_{\mu\nu}d\rangle$ and the
vacuum expectation values of four-quark operators.
Assuming the hypothesis of
vacuum dominance and/or unitarity symmetry to estimate the latter, it is
found that $\hat{B}=1.2\pm 0.1$

\vspace*{0.5cm}
\noindent

%\rule[.1in]{16.5cm}{.002in}

\newpage
\setcounter{page}{1}
\pagestyle{plain}

The $K_L$-$K_S$ mass difference  is rather sensitive to the mixing of the
t-quark and u- and c-quarks, so the analysis of the $K^0$-$\bar K^0$
system provides us with useful information about the values of mixing
angles
in the Kobayashi-Maskawa model and about the phase parameter $\delta$ of
$CP$-violation. The $K_L$-$K_S$ mass difference $\Delta m$ can be
presented
as a sum of the long-distance
dispersive contributions $\Delta m^L$ and the short-distance contributions
$\Delta m^{SH}$:
\begin{equation}
\Delta m =  \Delta m^L +  \Delta m^{SH}\, .
\end{equation}
$\Delta m^{SH}$ is related to the matrix element of the effective 
$\Delta S=2$
hamiltonian (see e.g. Ref.~\cite{Buras:1984ch}):
\begin{eqnarray}
\Delta m^{SH}& =& 2 {\rm Re} 
\langle \bar K^0| H_{eff}^{\Delta S =2}| K^0 \rangle\\ \nonumber
&=&(G_F^2/16\pi^2)F(x_j,\theta_j)(M_W^2/m_K)
\langle \bar K^0| \hat O| K^0 \rangle ,
\end{eqnarray}
where $\hat{O}=(\overline{s}_L\gamma_{\alpha}d_L)^2\alpha_s(\mu)^{-2/9}$
is
the renormalization-invariant operator of the hamiltonian, which arises in
the calculation of the well-known box diagram \cite{Gaillard:1974hs},
$G_F$ is the Fermi constant, and $M_{W}$ is the mass of the $W$-boson.
The function
$F(x_j,\theta_j)$
has the following form \cite{Inami:1981fz}:
\begin{equation}
F(x_j,\theta_j) ={\rm Re}[\lambda_c^2 S(x_c)\eta_1+\lambda_t^2 S(x_t)\eta_2+
2\lambda_c \lambda_t S(x_c,x_t)\eta_3]\, .
\end{equation}
Here $x_j=m_j^2/M_W^2$, $\lambda_i=V^{*}_{id}V_{is}$ ($V_{ij}$ is the
Kobayashi--Maskawa matrix), and the functions $S$ are defined as
\begin{eqnarray}
S(x)
  &=&x
\left[\frac{1}{4}+\frac{9}{4}(1-x)^{-1}-\frac{3}{2}(1-x)^{-2}\right]
+\frac{3}{2}[x^3/(1-x)^{3}]\ln x \, , \\ \nonumber
S(x_i,x_j) &= & x_i x_j
  \left\{\left[\frac{1}{4}+\frac{3}{2}(1-x_i)^{-1}\right.
-\frac{3}{4}(1-x_i)^{-2}\right] \\ \nonumber
&&\left.
\times \ln x_i/(x_i-x_j)+(i\leftrightarrow  j)
-\frac{3}{4}(1-x_i)^{-1}(1-x_j)^{-1}\right\}\, .
\end{eqnarray}
The coefficients $\eta_{i}$ make allowance for the strong-interaction
corrections in the leading logarithmic approximation. For
$\Lambda_{\overline{MS}}=100$ MeV they have 
the following numerical values: $\eta_1=0.7$, $\eta_2=0.6$,
and $\eta_3=0.4$ \cite{Gilman:1979bc}. The coefficients $\lambda_i$ are
connected with the mixing angles in the following way:
\begin{equation}
{\rm Re} \lambda_c \simeq s_1^2  c_1^2, \quad
{\rm Re} \lambda_t \simeq s_1^2  c_1^2 s_2^4, \quad
{\rm Re} \lambda_c \lambda_t \simeq s_1^2  c_1^2 s_2^2
\end{equation}
where $s_j=\sin~\theta_j$ and $c_j=\cos~\theta_j$, and the experimentally
acceptable hypothesis that $\sin~\theta_3\simeq 0$ and $\sin~\theta_2
\ll \sin~\theta_1$ is used. The mixing angles have the following numerical
values: $s_1=0.229\pm 0.006$, $c_1=0.9735\pm0.0015$, 
and $0.016< s_2< 0.095$ \cite{Wilkinson}.

As follows from eq.~(2), in order to calculate the short-distance
contribution to the $K_L$-$K_S$ mass difference, it is necessary to find
the value of the matrix element
\be
\langle \bar K^0|\hat O|K^0 \rangle=\frac{2}{3}f_K^2 m_K^2 \hat B
\ee
which is usually expressed through the dimensionless parameter $B$.

As for the long-distance contributions, they have been estimated
using different phenomenological approaches. For instance, the authors of
refs.~\cite{Wolfenstein:1979wi,Hill:1980ux} propose to estimate these
contributions by inserting the low-lying states between two 
$\Delta S=1$ weak
non-leptonic hamiltonians. More precise estimates have been obtained in
refs.~\cite{Cea:1985gj}-\cite{Pennington:1985it};
the corresponding values of the
parameter $D=\Delta m^L/\Delta m$ are $D=0.10\pm0.41$ \cite{Cea:1985gj},
$D=0.33\pm0.37$ \cite{Bigi:1984xh}, and $D=0.46\pm0.13$
\cite{Pennington:1985it}.

A lot of attention has been paid recently to estimating the parameter
$B$. The first such estimate -- $B=1$ -- has been obtained in
ref.~\cite{Gaillard:1974hs} using the vacuum dominance approximation.
Substituting this value into formula (2) and taking into account only the
contribution of the c-quark with the mass $m_c=1.3-1.5$ GeV, we find that
the short-distance contribution is about 40-45 $\%$ of the total value of
$\Delta m$. The t-quark contribution is small: for $m_t=40$ GeV and
$\tau_B \geq 10^{-12}$s it does not exceed 2$\%$ of $\Delta m$. Thus at
$B=1$ the short-distance contribution does not saturate the experimental
$K_L$-$K_S$ mass difference, which indicates that the long-distance
effects may be also important. Unfortunately, the current estimates
of $B$ depend considerably on both the particular model used in
the calculation and the values of various parameters involved in the model
\cite{Gaillard:1974hs},\cite{Shrock:1979dm}-\cite{Cea:1984xz}.
Thus at present it is not clear what is the real value of this
important parameter.

This paper presents a new estimation of the value
of $B$ within the QCD finite energy sum rules (FESR) method, which has
demonstrated
its efficiency in studying the properties of low-lying hadronic resonances
\cite{Chetyrkin:1978ta,Krasnikov:1982vw}.

Within this approach, the problem to be solved is reduced to computing the
following three-point correlator \cite{Nesterenko:1982fw}:
\bea
&&T_{\mu\nu}(p,q) =i^2\int dx dy \exp(ipx-iqy)
\times \langle Tj^5_\mu(x)\hat O(y) j^5_\nu(0) \rangle_0\\ \nonumber
&&
p_\mu q_\nu T(p^2, (p-q)^2,q^2)+ {\rm other~structures}
\eea
at small $q$ and large (euclidean) $p$: $|p^2|\geq 1$ GeV$^2$.
Here $j_{\mu}^{5}=\overline{d}\gamma_{\mu}\gamma^{5}s$ is the
interpolating field of the $K^0$ meson
\begin{displaymath}
\langle 0|j_{\mu}^{5}(0)|K^0(p)\rangle=if_{\rm K}p_{\mu}~~,
~~f_{\rm K}=1.17f_{\pi}=156~{\rm MeV}
\end{displaymath}
The matrix element (6) is related to the amplitude
$T(t)=T(-t,-t,0)$ by means of the dispersion
relation in the variable $t$:
\begin{eqnarray}
T(t) &=&  \int ds\frac{\rho(s)}{s+t}-{\rm subtractions} \\ \nonumber
&=&f_K^2\frac{\langle \bar K^0|\hat O|K^0 \rangle}{(t+m_K^2)^2}+
\frac{A}{t+m_K^2}+\ldots\, ,
\end{eqnarray}
where the single-pole contribution corresponds to the transition form
factor
of the $K^0$ meson, and the dots stand for the contributions of higher
states.

In the vacuum dominance approximation $T_{\mu,\nu}$ assumes the form
\begin{displaymath}
T_{\mu\nu}^{VD} = 
\frac{2}{3}\Pi_{\mu\alpha}(p)\Pi_{\nu\alpha}(q-p)\, ,
\end{displaymath}
\begin{displaymath}
\Pi_{\mu\alpha}(p)=i\int dx \exp(ipx)
\langle T(j^5_\mu(x) \bar s_L(0)\g_\al d_L(0)) \rangle_0\, ,
\end{displaymath}
and the resulting value of $B$ proves to be $B^{VD}=1$. Thus, there
remains to be computed
only the function $\Delta_{\mu\nu}=T_{\mu\nu}-T_{\mu\nu}^{VD}$, which is
responsible  for all the departures from the vacuum dominance
prediction for $B$.
In other words, within the sum rule approach we will calculate the value
$B-1$.

In the kinematical region we are interested in, there are three distinct
contributions to $T_{\mu\nu}$, viz: the perturbation theory (PT)
contribution; the power corrections due to the non-zero vacuum expectation
values ({\rm VEV}'s)
of local operators \cite{Shifman:1979bx}; and finally, the power
corrections
proportional to two-point correlators depending on the momentum $q$
(or, in other words, the {\rm VEV}'s  of bilocal operators
\cite{Balitsky:1983xk}). These bilocal power corrections are due to the
fact that one of the external momenta ($q$) is small, and thus, the
contribution from the region of large $y\sim 1/|q|$ in the integral (7)
cannot, generally speaking, be disregarded.

It is easy to check that the leading PT contribution to $T_{\mu\nu}$ is to
be completely assigned to $T_{\mu\nu}^{\rm VD}$. Thus, neglecting all
the non-leading PT corrections we find that
$\Delta_{\mu\nu}|_{p^2\rightarrow-\infty} 
= \Delta_{\mu\nu}^{\rm B}+\Delta_{\mu\nu}^{\rm L}$, 
with  $\Delta_{\mu\nu}^{\rm L}$ ($\Delta_{\mu\nu}^{\rm B}$)
standing for the terms proportional to the {\rm VEV}'s of (bi)-local
operators.

The account of local operators with dimension $\leq 6$ leads to the
following
result for the function  $\Delta_{\mu\nu}^{\rm L}$:
\begin{eqnarray}
\Delta_{\mu\nu}^{\rm L} &=&p^\mu q^\nu (-5 (pq)
\langle \al_s G^2 \rangle /{192\pi^3}-
4\langle \bar d s \bar d s \rangle
 \\ \nonumber
&&-4\langle \bar d  d\bar s s   \rangle +2\langle \bar s  s\bar s s
\rangle
+2\langle \bar d  d\bar d d   \rangle  \\ \nonumber
&&+m_s \langle g\bar d G_{\mu\nu} \sigma_{\mu\nu}d\rangle/{24\pi^2})
{p^{-2}(p-q)^{-2}} \\ \nonumber
&&+{\rm other~structures}
\end{eqnarray}
where the designation $\langle\overline{d}s\overline{s}d\rangle$ stands
for $\langle\overline{d}_{\rm L}\gamma_{\alpha}s_
{\rm L}\overline{s}_{\rm L}
\gamma_{\alpha}d_{\rm L}\rangle$ and so on;
$G^2=G_{\mu\nu}^aG_{\mu\nu}^a$;
$G_{\mu\nu}=G_{\mu\nu}^{a}t^{a}$; $tr(t^at^b)=\frac{1}{2}\delta^{ab}$;
and $\sigma_{\mu\nu}=\frac{1}{2}i[\gamma_{\mu},\gamma_{\nu}]_{-}$.

To estimate the contributions due to bilocal operators one needs to
construct
the Wilson expansion for the $T$-product $i\int dx \exp(ipx)
T(j_{\mu}^{5}(x)j_{\nu}^{5}(0))$ at large euclidean $p$ (or, equivalently,
at
small $x$). A straightforward calculation shows that in the leading order
in $\alpha_s$, the bilocal contribution to the tensor structure
$p_{\mu}q_{\nu}$ is suppressed by the factor $p^{-6}$ and, thus, can be
neglected within our approximation.

To proceed, the combination of four-quark operators appearing in (9)
proves to transform as a component of an $SU^f(3)$ 27-plet. This means
that the corresponding {VEV} is at least of second order in the
unitary-symmetry breaking parameter. Moreover, the {VEV} of each
four-quark operator in  (9) vanishes if the vacuum saturation procedure
\cite{Shifman:1979bx} is to work. Thus, the corresponding contribution to
$\Delta_{\mu\nu}^{\rm L}$ is ``doubly'' forbidden and can be safely
neglected. On the other hand, a straightforward estimation shows that
should both of these suppressions be absent in the next-to-leading
approximation, the
corresponding contribution to $\Delta B$ might be uncomfortably large
despite
the loop suppression factor $\sim \alpha_s/\pi$. To clarify the situation
we have calculated the $\alpha_s$-corrections to the coefficient functions
of the
four-quark terms in expansion (9) and have found that the appearing extra
terms are die out after applying the vacuum saturation.

In order to extract information on the value of $B$ we employ the FESR
technique to the function $T(t)$ multiplied by $(m_K^2+t)$ to nullify the
effect of the (unknown) single-pole contribution to the RHS of (8).
The final sum rule has the form
\begin{eqnarray}
B - 1&=& (\frac{2}{3}  f_K^4 m_K^2)^{-1}\int_0^{s_0} \rho^{th}(s) (s +
m_K^2) ds
\\ \nonumber
&=& m_s
\langle g\bar d\sigma_{\mu\nu}G_{\mu\nu} d \rangle/ 16\pi^2 m_K^2 f_K^4
\end{eqnarray}
where
\begin{displaymath}
\rho^{th}(s)=(2\pi i)^{-1}{\rm lim}|_{\epsilon\rightarrow 0}
\bigg[T(-s-i\epsilon,-s-i\epsilon,0)-T(-s+i\epsilon,-s+i\epsilon,0)\bigg],
\end{displaymath}
and $s_0=1.2$ GeV$^2$ is the duality interval of the $K^0$ meson
\cite{Krasnikov:1982vw}.

Up to now we have neglected all the effects due to the non-zero
anomalous dimensions of the operators under consideration.
In the leading log approximation
the account of these effects is carried out without any difficulty.
We define the renormalization-group invariant quantity
$\hat{B}=B(\mu)[\alpha_s(\mu)]^{-2/9}$ and make use of
the renormalization group technique \cite{BSh} to rewrite eq.~(10) as
\begin{displaymath}
\hat{B}=(1+ m_s
\langle g\bar d\sigma_{\mu\nu}G_{\mu\nu} d \rangle/ 16\pi^2 m_K^2
f_K^4|_{s_0})
[\al_s(s_0)]^{-2/9}\, .
\end{displaymath}

To estimate the numerical value of $B$, let us reduce the matrix
element $\langle g\overline{d}G_{\mu\nu}\sigma_{\mu\nu}d\rangle$
according to the relation $\langle
g\overline{d}G_{\mu\nu}\sigma_{\mu\nu}d\rangle$=
$m_0^2\langle\overline{d}d\rangle$, $m_0^2=0.8\pm0.4$ GeV$^2$
\cite{Belyaev:1982sa} and use PCAC relation
\begin{displaymath}
(m_u+m_d)\langle\overline{d}d\rangle = -\frac{1}{2}f_{\pi}^2m_{\pi}^2~,
\end{displaymath}
whence
\begin{displaymath}
B -1 = -(f_{\pi}^2 m_{\pi}^2/64 \pi^2 m_K^2 f_K^4)2m_s m_0^2/(m_u+m_d).
\end{displaymath}
The quark mass ratio is known with a relatively high accuracy
\cite{Gasser:1982ap}: $2m_s/(m_u+m_d)=25.0 \pm 2.5$. Finally, choosing
$\Lambda=100$ MeV we find
\begin{equation}
\hat{B}=1.2 \pm 0.1~~.
\end{equation}

Prediction (11) is in agreement with the calculation of $B$ within the
lattice approach \cite{Cabibbo:1984xa}, and also with the constraints
from above $B< 2.0 \pm 0.5$ \cite{Guberina:1983sz} and the result
$B=0.9$--$1.2$ \cite{Cea:1984xz}, which were obtained by the method
employing the dispersion representation for two-point Green functions.
At the same time, our result is about three times as large as the
value of $B\approx 0.3$--$0.4$ computed in the bag model
\cite{Colic:1983yb} and by using PCAC hypothesis and the $SU^{f}(3)$
symmetry in ref.~\cite{Donoghue:1982cq}.

To conclude, we have considered the problem of calculating the matrix
element (6) within QCD FESR approach. We have found that the bulk of
contributions to the relevant Green function $T_{\mu\nu}$ keeps within
vacuum dominance approximation, and thus, their effect on the value of
$B$ can be taken into account even without  any calculations! In our
opinion, this is the principal advantage of our method which most of
the other approaches apparently do not have. It has allowed us to
compute the parameter $B$ with a fairly high accuracy in spite of the
noticeable uncertainty involved in the determination of the
non-factorizable contributions to $T_{\mu\nu}$. Indeed, we have shown
that in the leading order in $\alpha_s$, the deviation of the actual
value of $B$ from its vacuum dominance estimate $B^{VD}=1$ is within the
following limits: $0.1 \leq (\hat{B}-1)\leq 0.3$, where the main
uncertainty is due to our poor knowledge of the matrix element
$\langle g\overline{d}G_{\mu\nu}\sigma_{\mu\nu}d\rangle$. However, the
overall magnitude of $\hat{B}-1$ proves to be small, which gives the
above margins for $\hat{B}$.

The authors are grateful to V.A. Matveev and A.N. Tavkhelidze for their
constant help and support. We thank N.V. Krasnikov, Z. Maki, V.A. Rubakov
and
M.E. Shaposhnikov for numerous discussions. One of us (K.G.Ch.) is
grateful
to P. Langacker for a useful discussion of the problem.

{\it Note added.}  After this work has been issued as a preprint, we
have learnt of another calculation of $B$ with the result
$\hat{B}=(0.33\pm 0.09) [\alpha_s(\mu^2)]^{2/9}$, which is quite
different from ours. In fact, the authors of ref.~\cite{Pich:1985ab}
have combined the effective chiral Lagrangian approach with the finite
energy sum rules method and calculated the parameter
$B(t=4m_{K}^2)=\langle 0| \hat{O}| K^0 K^0 \rangle/(2 f_K^2 m_K^2/3)$
rather than the parameter $B$ defined as in the
formula (6).

To our mind, one of the possible reasons for this disagreement could
be the noticeable variation of the function $B(t)$ between $t=4m_K^2$
and $t=0$ (within the chiral perturbation theory the quantity $4m_K^2$
can hardly be considered small in any way). It should be stressed that
our approach meets no such problem since from the very beginning we are
dealing with $B(0)\equiv B$.

Another possible explanation to the discrepancy could be the bad
convergance of the power correction series for the correlator examined in
ref.~\cite{Pich:1985ab}. Let us recall that in this work the account
of the two first power corrections have led to the decrease of the
result by five times, while in our analysis the power corrections to
the (known beforehand) factorizable contribution does not exceed
20$\%$ of the latter.

Leaving aside these technical subtleties, we feel that the main
advantage of our approach is the exact account of the factorizable
contributions and that the only chance to vary the result (11), say,
by a factor of 2 or 3 is to have a strong violation of the vacuum
saturation hypothesis for four-quark operators.

At present the problem of estimating the accuracy of the latter
hypothesis and the possibility of the exact account of factorizable
terms within our approach are under consideration.

\begin{center}
\large \bf Comments
\end{center}

Since the time the paper was published (1986) the need in an accurate
numerical value for the $\hat{B}_K$ has considerably raced up.
Precise experimental data on $K^0-\bar{K}^0$ system have appeared that
allowed to reliably determine $\epsilon'/\epsilon$. The
program of computing the next-to-leading correction to weak
Hamiltonian in the effective theory approach has been completed
\cite{H_weak:NLO}. However, no significant improvement in calculation
of the matrix element has been achieved. The reason is clear: it is a
non-PT problem. As such it is now actively discussed within lattice
approach but the corresponding results are still far from being perfect.

Our  calculation was made within the standard continuous QCD and the
QCD Sum Rule Approach. The method we used factors out completely (and
thereby takes into account) the factorizable contribition to the
matrix element in all orders of PT and OPE expansion. The remaining
non-factorizable part happens to be small in comparison to
factorizable term, the latter being accounted exactly in our set-up.

The method is open to improvement: both high order PT corrections 
and next order power
corrections in addition to those computed in the paper
can be added. Note that PT corrections are really
mandatory to compute to match the available NLO coefficient functions
(see, e.g. \cite{Burasetal}).

Having in mind a solid theoretical basis of our calculation we have just
updated the prediction for $\hat B_K$ by taking into account a
significantly changed value for $\Lambda_{\rm\overline{MS}}^{(f=3)}$ 
from 100 MeV to the value of over 400 MeV, which corresponds 
to the world average value $\alpha_s(M_Z)\approx 0.118$. The updated version
of eq.~(11) with $\al_s(1.2~{\rm GeV}^2)=0.69$, estimated  in the 
leading order, is :
\[ 
\hat B = 1.0 \pm 0.1
\]
Note that the bulk of the change of the parameter $\hat B$
is due to the normalization (the change in the factor 
$\al_s(1.2~{\rm GeV}^2)^{-2/9}$) while other
parameters of the calculations entering eq. (10)
did not change much during last 15 years.

\end{document}